\documentclass[%
 reprint,
 amsmath,amssymb,
 aps,
]{revtex4-2}

\usepackage[table]{xcolor}
\usepackage{graphicx}
\usepackage{tikz}
\usetikzlibrary{arrows}
\usetikzlibrary{arrows.meta}
\usetikzlibrary{decorations.pathmorphing}
\definecolor{dark-green}{RGB}{12,84,0}
\definecolor{dark-red}{RGB}{143,34,38} 
\definecolor{dark-blue}{RGB}{0,0,120}
\usepackage{pgfplots}
\pgfplotsset{width=14cm,compat=1.15}
\usepgfplotslibrary{patchplots} 

\usepackage{adjustbox}
\usepackage{cancel}
\usepackage{multirow}
\usepackage{amssymb}
\usepackage{amsmath}
\usepackage{verbatim}
\usepackage[subrefformat=parens,labelformat=parens,caption=false]{subfig}
\usepackage{soul}

\usepackage{wasysym}

\usepackage[sort&compress]{natbib}
\usepackage{hyperref}
\hypersetup{
    colorlinks=true,       
    linkcolor=black,        
    urlcolor=black,       
    citecolor=black, 
}

\usepackage{subfig}

\usepackage{devanagari}

\newcommand{\C}{\mathcal{C}}

\begin{document}
%\title{Emergent Behaviors of Active Particles on Deformable Surfaces: Clustering, Mixing, and Localization }
\title{Spatial organisation of multiple species of active particles interacting with an interface}
\author{Love Grover}
\email{ph17047@iisermohali.ac.in}
\affiliation{Indian Institute of Science Education and Research, Mohali, India}

\author{Rajeev Kapri}
\email{rkapri@iisermohali.ac.in}
\affiliation{Indian Institute of Science Education and Research, Mohali, India}

\author{Abhishek Chaudhuri}
\email{abhishek@iisermohali.ac.in}
\affiliation{Indian Institute of Science Education and Research, Mohali, India} 
 
\date{\today}

\begin{abstract}
    We investigate the steady-state organisation of active particles residing on an interface. Particle activity induces interface deformations, while the local shape of the interface guides particle movement. We consider multiple species of particles which can locally pull on the interface or push it. This coupled system exhibits a wide variety of behaviours, including clustering, anti-clustering, diffusion, mixing, demixing, and localisation. Our findings suggest that one can control surface properties by strategically adding or removing specific particle types.  Furthermore, by adjusting particle activity levels, we can selectively disperse particle types, enabling precise manipulation of surface movement and geometry.
\end{abstract}

\maketitle

\graphicspath{{plots/},{data/plotcdf}}

\section*{Introduction}
The cell membrane, a vital component orchestrating various cellular functions such as migration, motility, endocytosis, exocytosis, and cell adhesion, is inherently dynamic~\cite{jarsch2016membrane,mcmahon2005membrane,mcmahon2015membrane}. Understanding the complex dynamics of such processes, particularly the shape fluctuations in the cell membrane and the advection of lipids and proteins by these fluctuations to form clusters on the membrane, is critical~\cite{kozlov2014mechanisms,zimmerberg2006proteins,gov2018guided}. Several components contribute to the membrane's shape-changing dynamics, including the membrane's local lipid composition, the cell cytoskeleton, the nature of the membrane-bound proteins and the activity of membrane channels and pumps~\cite{ramaswamy1999nonequilibrium,ramaswamy2000nonequilibrium,gov2004membrane,kozlov2014mechanisms,liu2010mechanochemical,baumgart2011thermodynamics,simunovic2015physics,veksler2007phase,gov2006dynamics,sorre2012nature,sharma2004nanoscale,goswami2008nanoclusters,gowrishankar2012active,koster2016actomyosin,dawson2006bar}. In most cellular processes, a combination of these components determines the shape fluctuations and the clustering of lipids and proteins~\cite{prevost2015irsp53,galic2012external,suetsugu2014dynamic}. {\em In vitro} studies of lipid membranes with variations in lipid composition and embedded curved membrane proteins have been successfully utilised to understand the importance of each of these components. The membrane proteins can induce local membrane curvature. At the same time, membranes influence the dynamics of these proteins, highlighting a reciprocal relationship between membrane architecture and protein dynamics. The actomyosin complex below the cell membrane further gives rise to active forces on the membrane, which influences these dynamics significantly. These forces could be due to the local polymerisation of the actin biopolymers and/or the contractile forces due to the myosin motors. Theoretical studies have explored the coupling between the membrane-bound proteins and the cytoskeletal forces and their effect on membrane shape and dynamics~\cite{leibler1986curvature,reynwar2007aggregation,shlomovitz2007membrane,chaudhuri2011spatiotemporal}. 

Statistical models built on the generic features observed in real systems can help understand much of the rich physics in such systems. Therefore, it is prudent to ask if the active processes involved in shaping membrane fluctuations and the spatiotemporal organisation of the membrane proteins can be incorporated into a simpler model with features that mimic the complex setting. Recently, Cagnetta et al ~\cite{cagnetta2018active,cagnetta2019statistical,cagnetta2022universal} built on a minimal statistical model of passive particles on an interface~\cite{das2000particles,das2001fluctuation,nagar2005passive,nagar2006strong,gopalakrishnan2004dynamics} to look at the patterns formed by active non-interacting particles on such an interface. Essentially, the cell membrane is modelled as a one-dimensional interface with active components embedded in it. These active components are assumed to mediate the various local forces by acting as ``pullers", which pull the interface upward, and ``pushers", which push the interface downward. This model probed the coupling between pullers and the interface, akin to the coupling between growth-promoting membrane-bound proteins and the cell membrane. They found that particles organise in microclusters and drive the interface, mimicking a migrating cell membrane. In a later study, Bisht and Barma~\cite{bisht2019interface} studied the interaction between a single active inclusion (pusher or puller) and the interface. They found that a puller moves superdiffusively in the transverse direction while pulling the immobile interface upward. 
A pusher moves subdiffusively, leading to a separation of timescales in pusher motion and interface response. 

A cell membrane, however, is subjected to simultaneous pushing and pulling forces. Therefore, it is natural to ask about the patterning of particles on the interface if it is under the influence of both species of active particles - pushers and pullers. Chakraborty et. al.~\cite{chakraborty2017static,chakraborty2017dynamic} studied a system where a fluctuating energy landscape advects two species of particles, one lighter (L) and the other heavier. In this Light-Heavy (LH) model, the particles move along the potential gradient, but each type of particle alters the local landscape differently. This interplay between particles and landscape leads to various ordered states.  Depending on how each particle type affects the landscape, the landscape can either be fully ordered with distinct regions or a mix of ordered and disordered areas. 

\begin{figure*}[t]
\centering
    \includegraphics[width=\textwidth]{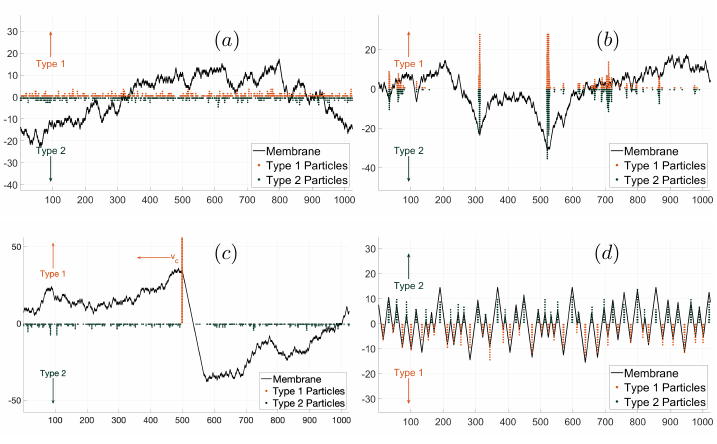}
    \caption{Snapshots of different membrane configurations and particle distributions during the evolution of a two-particle system across all four cases: (a) $\{+,-,-,+\}$ not significantly different spatial organization from the single-particle case $\{-,+\}$, consisting of two opposite particles; (b) $\{+,+,-,+\}$, showing an almost equal distribution of both particle types; (c) $\{-,-,-,+\}$, forming a large Type 1 cluster; (d) $\{+,+,-,-\}$, representing a locked-in state. Type 1 and Type 2 particles are plotted along the positive and negative directions of the y-axis, as indicated in the figures, for better clarity. In (c), $v_c$ is the velocity of the cluster, which has a direction opposite to the smooth slope.}
   ~\label{fig:anime}
\end{figure*}

In this paper, we study a minimal statistical physics model which assimilates some of the principles of the LH model, considering multiple species of active particles on an interface without excluded volume interactions. This paper focuses on the spatial patterning/organisation of particles on the interface. Firstly, we analyse single-type active growth-promoting particles (pullers) on the interface.  Two parameters, denoted as the `activity' parameter $\beta$ and the `gravity' parameter $\gamma$, regulate the dynamics of this system. Specifically, the flow of particles is influenced by $\gamma$ and the local shape of the interface.
On the other hand, the interface behaviour is dictated by $\beta$ and the distribution of particles on it. We look at the normalised two-point density-density correlation as a function of these parameters. The collective patterns show both correlated and anti-correlated behaviour. Next, we analyse the particle organisation when two species - pushers and pullers - are present in the interface. Even within a controlled parameter setting, this system shows many possible states, such as clustering, anti-clustering, diffusive, mixing, demixing, and localisation, as some of them can be clearly seen in Fig \ref{fig:anime}.

\section{Model}~\label{sec:model}

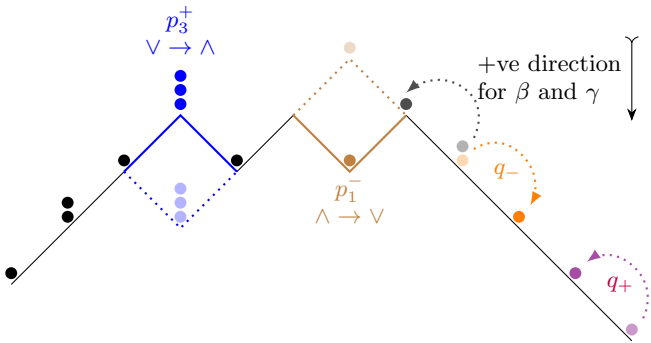
\begin{figure}
  \begin{tikzpicture}[scale=0.75]
      \definecolor{tempcolor}{rgb}{1,0.73,0.73} 
      %new
      \node[text width=2.3cm] at (9.8,3.65) {+ve direction\\ for $\beta$ and $\gamma$};
      \node[color=blue] at (3,4.7) {$p_3^+$};
      \node[color=blue] at (3,4.2) {$\vee\rightarrow\wedge$};
      \node[color=brown] at (6,1.7) {$p_1^-$};
      \node[color=brown] at (6,1.2) {$\wedge\rightarrow\vee$};
      \node[color=orange] at (8.8,2) {$q_-$};
      \node[color=purple] at (10.8,0) {$q_+$};
      
      %end new
      \draw (0,0) -- (1,1) -- (2,2); 
      \draw (4,2) -- (5,3); 
      \draw (7,3) -- (8,2) -- (9,1) -- (10, 0) -- (11, -1); 
      \draw[brown,thick] (5,3) -- (6,2) -- (7,3); 
      \draw[brown, dotted, thick] (5,3) -- (6,4) -- (7,3);
      \draw[>-Stealth] (11,4.4) -- (11,2.9);
      \draw[blue,thick] (2,2) -- (3,3) -- (4,2); 
      \draw[blue, dotted,thick] (2,2) -- (3,1) -- (4,2); 
      \fill[black](0,0.20) circle (3pt); 
      \fill[black](1,1.20) circle (3pt); 
      \fill[black] (1,1.45) circle (3pt);  
      \fill[black] (2,2.20) circle (3pt); 
      \fill[black] (4,2.20) circle (3pt);
      \fill[blue] (3,3.20) circle (3pt);
      \fill[blue] (3,3.45) circle (3pt); 
      \fill[blue] (3,3.70) circle (3pt); 
      \fill[blue!30](3,1.20) circle (3pt);
      \fill[blue!30] (3,1.45) circle (3pt); 
      \fill[blue!30] (3,1.70) circle (3pt);
      \fill[brown] (6,2.20) circle (3pt);
      \fill[brown!30] (6,4.20) circle (3pt);
      \fill[black!70] (7,3.20) circle (3pt); 
      \fill[black!30](8,2.45) circle (3pt); 
      \fill[orange!30] (8,2.20) circle (3pt); 
      \fill[orange] (9,1.20) circle (3pt); 
      \fill[violet!40] (11,-0.80) circle (3pt); 
      \fill[violet!70] (10,0.20) circle (3pt); 
      \draw[-latex,orange,thick,dotted] plot [smooth,tension=1.5] coordinates {(8.2,2.4) (9.05,2.3) (9.2,1.4)}; 
      \draw[-latex,black!70,thick,dotted] plot [smooth, tension=1.5]coordinates{(8.2,2.6) (8.0,3.5) (7.0,3.4)};
      \draw[-latex,violet!70,thick,dotted] plot [smooth, tension=1.5]coordinates{(11.2,-0.6) (11.05,0.3) (10.2,0.4)};
  \end{tikzpicture}
  \caption{
A schematic representation of the proposed model illustrating the interface profile updates and the dynamics of the active particles. The blue region shows the transformation of a valley (depicted as dots) into a hill (solid blue lines). In contrast, the brown region shows the transformation of a hill (depicted as dots) into a valley (solid brown lines). The downhill (uphill) motion of particles along the interface is depicted in orange (violet). Specific situations with their corresponding probabilities are also shown in the figure. Positive $\beta$ and $\gamma$ corresponds to a higher likelihood of a hill turning into a valley and a particle on an average moving down the slope respectively. This direction is also shown in the right top corner.}\label{fig:chain}
\end{figure}

Our model consists of $N$ particles on a one-dimensional lattice of $L$ sites, labelled $i = 1,2, ..., L$, with spacing $a = 1$ between the sites and with periodic boundary conditions~\cite{das2000particles,cagnetta2018active,bisht2019interface}. The particles do not have hard-core constraints as in Ref.~\cite{cagnetta2018active}. Therefore, there can be more than one particle at a site, with $s_i$ denoting the occupancy of a site $i$. Every site in the lattice is linked with a bond placed at half-integer sites. Each such link, indexed as $i- \frac{1}{2}$, has a slope $\tau_{i-\frac{1}{2}}=\pm 1$. Therefore, the bonds/links represent discrete interface elements with two possible orientations denoted by $\backslash$ or $/$. For example, using two slopes around $i$th site viz., $\{\tau_{i-\frac{1}{2}},\tau_{i+\frac{1}{2}}\}$, we get 4 possible shapes including hills ($\{+1,-1\}=\wedge$) and valleys  ($\{-1,+1\}=\vee$) as described in Fig.~\ref{fig:chain}. To study the dynamics of the hills and valleys, we introduce the relative height profile ${h_i}$ at every site with $h_i = \sum_{1\leq j \leq i}\tau_{j - 1/2}$. The instantaneous average height is given by $\langle h(t) \rangle = \frac{1}{L}\sum_{i=1}^L h_i(t)$, which fluctuates over time.

\vskip 0.2cm
\noindent
{\em Interface updates:}
An interface update involves randomly selecting a site and performing stochastic local single-step moves. The update is performed only if the chosen site is a local hill or a valley. Therefore, the only possible moves that change the interface configuration are a hill converted to a valley ($\wedge\rightarrow\vee$) and a valley converted to a hill ($\vee\rightarrow\wedge$). These two moves will have the same rate in the absence of particles on lattice sites, and the local dynamics are then of the Edwards-Wilkinson (EW) type~\cite{nagar2006strong}. However, the presence of active particles on the sites dictates the flipping rates of hills and valleys. When there are $n$ particles on the chosen site, the update of the local interface occurs with the rates $p_n^{+}$ for hill $\rightarrow$ valley and $p_n^{-}$ for valley $\rightarrow$ hill and are given as:
\begin{equation}
    p^\pm_n=u\frac{e^{\pm n\beta}}{e^{n\beta}+e^{-n\beta}}.
~\label{newactivity2}
\end{equation}
where we call $\beta$ as the {\em activity} parameter. For $n = 0$, $p^+_0 = p^-_0 = u/2$, which gives the EW dynamics as mentioned before. For $n \neq 0$, $\beta > 0$, gives $p^+_n > p^-_n$, implying a higher likelihood of a hill turning into a valley than the opposite. Similarly,  $\beta < 0$ gives $p^+_n < p^-_n$, which implies a higher likelihood of a valley turning into a hill. Eq.~\ref{newactivity2} can be easily generalised if we consider more than one type of active particle. For example, for two types of active particles with numbers $\{ n_1, n_2 \}$ with activities $\{\beta_1, \beta_2 \}$ at a site, the update rules are given by 
\begin{equation}
    p^\pm_{\{n_1,n_2\}}=u\frac{e^{\pm (n_1\beta_1 +  n_2\beta_2) } }{e^{(n_1\beta_1 + n_2 \beta_2)}+e^{-(n_1\beta_1 + n_2\beta_2)}}.
~\label{nactivity}
\end{equation}

\vskip 0.2cm
\noindent
{\em Particle update:}
A particle can move on the lattice to its neighbouring sites (left or right). The transition rates depend on the local curvature of the interface, i.e., $\nabla h=(h_{i+1}-h_{i-1})/a$, where $h_{i+1} \ (h_{i-1})$ is the height of the right (left) site. The particle can move left (or right) with rates $q^+$ (or $q^-$) given by:
    \begin{equation}
       q^\pm = q_0 \left(1\pm\frac{\gamma}{2}\nabla h\right).
    \end{equation}
Therefore, when a particle is on a hill or a valley, the transition rates $q^+=q^-=q_0$, since $\nabla h = 0$. For other instances, $\gamma$ determines the rates for the particles to move up or down a slope, and we call it the {\em gravity} parameter. For example, for a randomly selected particle residing on-site $\{-1,-1\}=\backslash$, the rate of transition to the left (or right) is proportional to $q^\pm=q_0(1\mp\gamma)$. Therefore, if $\gamma > 0$, $q^+ < q^-$, the particle on an average moves down the slope (downhill) and if $\gamma < 0$, $q^+ > q^-$, the particle on an average moves up the slope (uphill). 
When $\gamma = 0$, the particles do not feel the interface and move with equal rates $q_0$ in either direction. Therefore, in this situation, the particle motion is purely diffusive. 

We consider a lattice of $L = 4096$ sites with $N = 4096$ particles. 
One Monte Carlo (MC) step constitutes $L$ interface update attempts and $N$ particle update attempts and is the unit of time used for this problem. Time averaging is done after $2\times10^6$ steps with data collected every $4096$ step. We choose $u = 1$ and $q_0 = 0.5$.

\begin{table}[t]
  \caption{%
    Symmetry analysis of a system with a single particle type,
    representing signs of $\{\beta,\gamma\}$.  Due to reversal
    symmetry, the two mathematically possible configurations
    represent equivalent physical systems, leaving only two
    distinct cases.
  }
  \label{tab:1particle}
  \begin{ruledtabular}
    \begin{tabular}{cc}
      Cases       & Reversal    \\
      \colrule
      $\{+,-\}$   & $\{-,+\}$   \\
      $\{+,+\}$   & $\{-,-\}$   \\
    \end{tabular}
  \end{ruledtabular}
\end{table}

\section{Results And Analysis}~\label{results}

\subsection{One Type Of Particle}
In this situation, only one set of parameters $\{\beta, \gamma \}$ govern the dynamics of the system. We adopt the following convention: $\beta = +(-)$ represents the situation when the particle pushes (pulls) the interface down (up); $\gamma = +(-)$ represents that the particle is going downhill (uphill). For example, the set $\{+,+\}$ represents the situation where an active particle going downhill pushes the interface down. Naturally, there are four different possibilities, $( \{+,-\}, \{+, +\}, \{-,+\}, \{-,-\})$ (see Table~\ref{tab:1particle}). However, only the first two are distinct. The other two can be obtained by interchanging the $+$ with the $-$ sign and vice versa.

\begin{figure}[t]
\centering
    \includegraphics[width=\columnwidth]{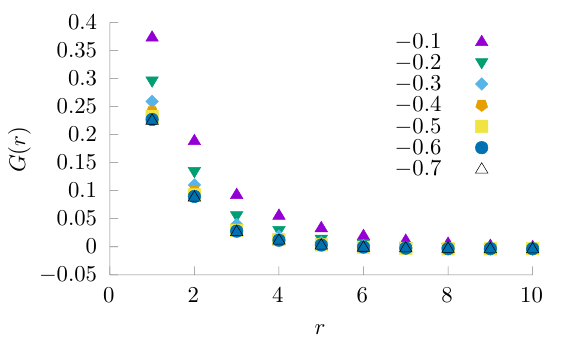}
    \caption{For a single kind of active particles on the interface, normalised two-point density correlations $G(r)$ for a fixed value of gravity parameter 
($\gamma = 0.8$) and different values of the activity parameter $\beta$. With higher activity, the correlation at lower length scales becomes smaller and eventually saturates. The saturation value depends on $\beta$ and $\gamma$. Here, $G(r) \sim 1/r^{\alpha}$ with $\alpha \sim 1$ for $r \rightarrow 0$}.
   ~\label{fig:corvsrfixedgamma}
   \end{figure}

  \begin{figure}[b]
  \centering
    \includegraphics[width=\columnwidth]{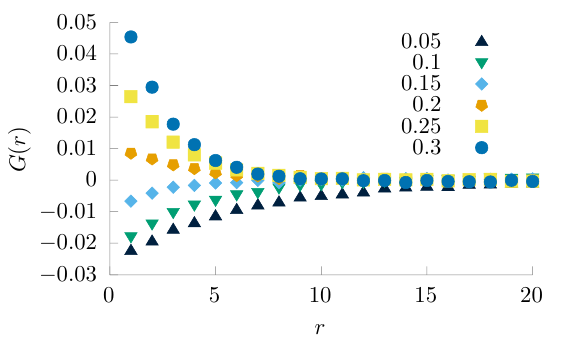}
    \caption{For a single kind of active particles on the interface, normalized two-point density correlations $G(r)$ for a fixed value of activity parameter
 ($\beta=-0.8$) and different values of the gravity parameter $\gamma$. With increasing $\gamma$, we see a cross-over from anti-correlated behaviour to correlated behaviour.}
    ~\label{fig:corvsrfixedbeta}
   \end{figure}

The main objective of our study was to explore cooperative effects when two species of particles are present. Therefore, for the single type of particle considered in this section, although there are two distinct possibilities of parameters, we choose the set $\{ -, +\}$ as a case study. Here, particles are of the ``puller" type, converting valleys into hills while themselves moving downhill. We have considered activity and gravity parameters in the range $-2 \le \beta \le 0$ and $0 \le \gamma \le 1$. To understand the system's behaviour for various parameter values, we calculate the two-point density-density correlation function, which gives valuable insight into the tendency of the particles to cluster~\cite{nagar2006strong}.

\begin{figure}[t]
    \includegraphics[width=1.0\linewidth]{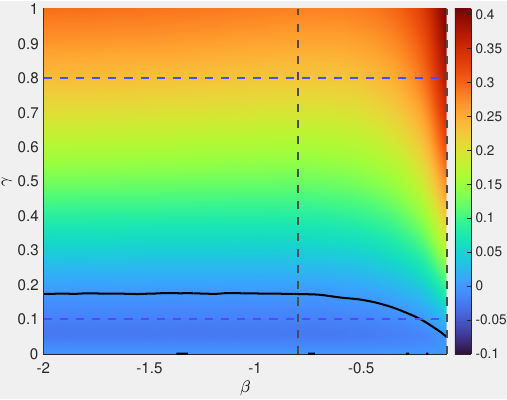}
    \caption{Heatmap of $G(r=1) \equiv G(1)$, showing three correlation regions: (a) positive-correlation ($>0$), (b) zero-correlation ($=0$, solid contour line), and (c) negative-correlation ($<0$). The dotted lines indicate the parameter variations used in  Figs.~\ref{fig:corvsrfixedgamma},\ref{fig:corvsrfixedbeta} and ~\ref{fig:cdf}. The saturation of $G(1)$ can be observed by inspecting the color gradients, where the contours become parallel to lines of constant $\gamma$.}
   ~\label{fig:heatmap} 
   \end{figure}

The normalised density correlation function is given in terms of the lattice site occupancy $s_i$ as:
\begin{equation}
    G(r) = \frac{ \langle s_i s_{i+r} \rangle - \langle s_i \rangle^2 }{ \langle s_i ^2 \rangle - \langle s_i \rangle^2}.
~\label{eq:Gr}
\end{equation}
where $i$ is the lattice site index and $r$ is the lattice distance. 
In Fig.~\ref{fig:corvsrfixedgamma}, we have plotted $G(r)$ as a function of $r$ for various activity ($\beta$) values for a fixed high value of $\gamma$. We observe that for all $\beta$ values, $G(r)$ decays to zero as $r$ increases. 
As $r \rightarrow 0$, $G(r) \sim 1/r^\alpha$, and $\alpha \rightarrow 1$ with increasing $|\beta|$. This behavior is very different from passive sliders on fluctuating surfaces~\cite{das2001fluctuation}. As already reported for this active interface \cite{cagnetta2018active}, density fluctuations do not scale with the system size. This $L$–independence indicates that there is no characteristic domain size growing with $L$, and the system does not display phase separation with well‐defined (sharp or fuzzy) interfaces. Instead, the fluctuations are significant at every scale.

Next, in Fig.~\ref{fig:corvsrfixedbeta}, we plot $G(r)$ as a function of $r$ for a fixed $\beta$ and various $\gamma$ values. For the trivial case of $\gamma = 0$, the particle can move left or right with an equal rate; the motion is diffusive and $G(r) = 0$ for all $r$. As $\gamma$ is increased from zero by a small value (e.g., $\gamma = 0.05$ and 0.1), we observe anti-correlation (i.e., $G(r)$ is negative) for smaller $r$. The anti-correlation decreases as we increase $\gamma$. For high values of $\gamma$, we again see the decay of correlation similar to that observed in Fig.~\ref{fig:corvsrfixedgamma}.  
This trend suggests a non-zero $\gamma$ value for which $G(r) \rightarrow 0$ for all $r$.

\begin{figure*}[t]
 \includegraphics[width=0.8\linewidth]{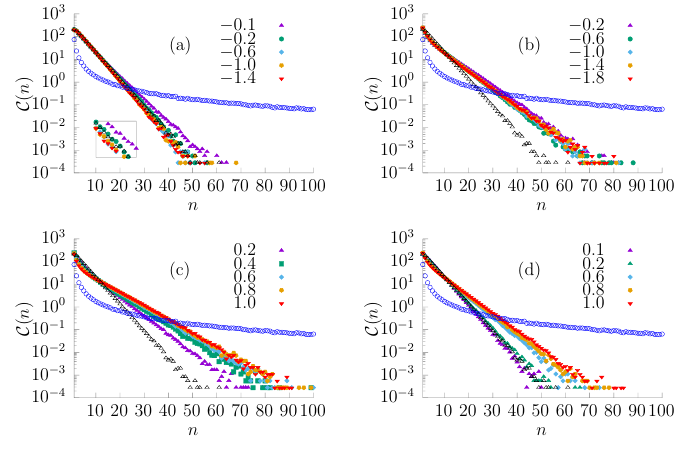}
    \caption{Cluster size distribution $\C(n)$ for active particles of a single type on the interface. The plots are shown for various parameter values $\beta$ and $\gamma$ in the log scale. (a) and (b) are for different activity parameter ($\beta$) values and at constant gravity parameter values, $\gamma = 0.1$ and $0.8$ respectively. (c) and (d) are for different gravity parameter ($\gamma$) values and at constant activity parameter values $\beta=-0.1$ and $-0.8$ respectively. In (a), the inset shows a magnified region of the plot. In all the plots the two special cases of $\gamma = 0, \beta = 0$ and $\gamma \neq 0, \beta = 0$ are shown as $\triangle$ and $\Circle$ respectively.  We consistently observe an exponential decay in the cluster distribution,  $C(n) \sim \exp{(-n/n^*)}$, with $n^* \sim 10$ for the active cases. In contrast, for passive systems without activity ($\gamma \neq 0, \beta = 0$), the distribution shows a power law decay with an exponent $\sim 1.5$, as previously reported in \cite{nagar:thesis}.}\label{fig:cdf}
\end{figure*}

To gain a broader understanding of the combined effects of $\beta$ and $\gamma$, we have shown a phase plot for $G(r = 1) \equiv G(1)$ in Fig.~\ref{fig:heatmap} in the range $-2 \le \beta < 0$, and $0 \le \gamma \le 1$. This reveals three distinct regions: (i) the anti-correlated region, (ii) the uncorrelated region (excluding the trivial $\gamma = 0$ case) and (iii) the positively correlated region. The significant decrease in correlation values by adding activity in the system suggests a competition between gravity-driven particle accumulation in valleys and activity-driven dispersion of the particles. It is natural to assume that gravity promotes particle clustering, while activity in the form of ``pullers" can destabilise clusters, particularly by transforming valleys (filled with particles) into hills. This competition is supported by the anti-correlations observed for lower $\gamma$ values. For $\beta \lesssim -0.3$, $G(1)$ saturates, and increasing activity has no further effect. The saturation value is dependent on $\beta$ and $\gamma$ as shown in Fig.~\ref{fig:heatmap}.  The saturation occurs because, beyond a certain activity, the break-up of the clusters is unaffected by it. At a fixed high value of $|\beta|$, if we increase $\gamma$, we enter the zero-correlation region. In the zero-correlation region, the effects of gravity and the activity effectively cancel each other, and therefore, the particle distribution is only affected by the stochasticity in the system. Increasing $\gamma$ further leads to positive correlations, as shown in Fig.~\ref{fig:heatmap}. $G(1)$ values are still small as activity suppresses the build-up of correlations in the system.

We also study the mean cluster size distribution function represented by $\C(n)$, where $n$ is the size of a cluster, for various $\beta$ and $\gamma$ values as shown in Fig.~\ref{fig:cdf}. Here, a cluster is defined as a contiguous group of particles such that empty sites on both ends bound the group. We have also plotted $\C(n)$ for two special cases in all the subplots as a benchmark. Firstly, when $\gamma = 0, \beta = 0$ (shown by symbol $\triangle$), the interface and particle dynamics are completely decoupled, and the particle motion is purely diffusive. In this case, the cluster distribution is exponential. Secondly, when $\gamma \neq 0, \beta = 0$ (shown by symbol $\Circle$), the particles slide downhill and form passive clusters as in Ref.~\cite{nagar2006strong}. The cluster distribution is a power-law type with a significant number of large clusters. In Fig.~\ref{fig:cdf}(a), we find that for low $\gamma$ values, increasing $|\beta|$ can push the distribution $\C(n)$ even lower than the purely diffusive case (as shown in the inset). This is the anti-clustering regime we discussed earlier. 

In Fig.~\ref{fig:cdf}(b), we plot $\C(n)$ for $\gamma = 0.8$. If we compare our result with the $\gamma \neq 0, \beta = 0$, we can conclude that increasing $|\beta|$ shifts the distribution from power law to exponential. In this case, the slopes are always higher than the purely diffusive system, unlike $\gamma = 0.1$ (Fig.~\ref{fig:cdf}(a)). This indicates that in all such cases, we have a larger number of clusters of bigger sizes. This feature has also been observed in Ref.~\cite{cagnetta2018active}.

So far, we have studied the case where we have fixed the gravity parameter $\gamma (= 0.1$,\,$0.8)$ and varied the activity parameter $\beta$, therefore moving horizontally from right to left in Fig.~\ref{fig:heatmap}. Now, we keep $\beta$ fixed in the same figure and move vertically from bottom to top. We have plotted $\C(n)$ vs $n$ for $\beta =-0.1$ and $\beta = -0.8$ in Figs.~\ref{fig:cdf}(c) and ~\ref{fig:cdf}(d), respectively. We can see that as $\gamma$ increases, we have more clusters of bigger size, thus showing more deviation from purely diffusive behaviour. In the $|\beta| = 0.8$ trends, we see that first $\C(n)$ falls below the purely diffusive case and then increases as $\gamma$ increases. This is reminiscent of the anti-clustering behaviour discussed in Fig.~\ref{fig:heatmap} and Fig.~\ref{fig:cdf}(a). 

In summary, the behaviour of the single-particle system reveals that increasing activity disrupts clustering, and the interplay between gravity and activity determines whether particles accumulate or disperse. We now investigate the more complex dynamics that can arise when two distinct species interact on the same interface. This additional complexity allows us to explore a wider range of particle behaviours and interface deformations.

\subsection{Two types of particles}

Since two particle species are considered here, we denote them as $1$ ($2$) representing first (second) type particles for each set. A particle set is represented as $\{\beta_1, \gamma_1, \beta_2, \gamma_2\}$ with $N/2$ particles of each type. For example, the set $\{+,+,-,+\}$ represents the situation where the type-1 particle goes downhill ($\gamma_1 = +$) and pushes the interface down ($\beta_1 = +$), while the type-2 particle goes downhill ($\gamma_2 = +$) but pulls the interface up ($\beta_2 = -$). Of the 16 different possibilities, only four types of systems can be made using two types of particles (see Table~\ref{tab:2particles}). The four distinct sets are $\{+,+,-,+\}$, $\{+,+,-,-\}$, $\{+,-,-,+\}$ and $\{-,-,-,+\}$. However, varying the parameters $\beta$ and $\gamma$ will lead to many possible scenarios even for these four distinct particle states. Therefore, we have fixed $\beta_{1,2} = \pm 2$ and $\gamma_{1,2} = \pm 1$ in our study. Although this restricts our parameter space, we show below that the organisation of the two species of particles on the interface shows a wide range of behaviours. We explore them by first looking at the cluster size distribution.

\begin{table}[b]
  \caption{%
    Analysis of system configurations with two particle types, representing signs of $\{\beta_1,\gamma_1,\beta_2,\gamma_2\}$, under symmetry operations (reversal and exchange). The 16 mathematically possible configurations reduce to four distinct system types, with the remaining cases being equivalent to these or representing trivial systems (all particles identical). The notation $C(a)$ indicates that after the symmetry operation, the configuration is identical to column $a$.}
  \label{tab:2particles}
  \begin{ruledtabular}
    \begin{tabular}{c c c c}
      & \multicolumn{3}{c}{Symmetries} \\
      \colrule
      Cases         & Reversal         & Exchange        & Both            \\
      \colrule
      $\{+,+,-,+\}$ & $\{-,-,+,-\}$   & $\{-,+,+,+\}$   & $\{+,-,-,-\}$   \\
      $\{+,+,-,-\}$ & $\{-,-,+,+\}$   & $C(2)$          & $C(1)$          \\
      $\{+,-,-,+\}$ & $\{-,+,+,-\}$   & $C(2)$          & $C(1)$          \\
      $\{-,-,-,+\}$ & $\{+,+,+,-\}$   & $\{-,+,-,-\}$   & $\{+,-,+,+\}$   \\
      \colrule
      \multicolumn{4}{c}{Trivial cases}                             \\
      \colrule
      $\{-,-,-,-\}$ & $\{+,+,+,+\}$   & $C(1)$          & $C(2)$          \\
      $\{+,-,+,-\}$ & $\{-,+,-,+\}$   & $C(1)$          & $C(2)$          \\
    \end{tabular}
  \end{ruledtabular}
\end{table}

\begin{figure*}[t]
\includegraphics[width=0.8\linewidth]{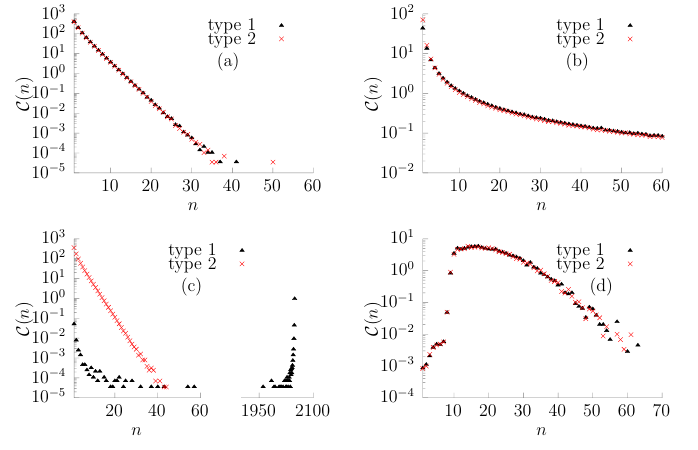}
    \caption{Steady state cluster size distribution $\C(n)$ for the case of two different types of active particles (labelled as type 1 and type 2) on the interface for four distinct combinations of signs of activity parameter ($\beta$) and gravity parameter ($\gamma$) values discussed in Table 2. The magnitude of parameters is taken as $|\beta| = 2$ and $|\gamma| = 1$ for all four cases. (a) $\{+,-,-,+\}$; (b) $\{+,+,-,+\}$; (c) $\{-,-,-,+\}$; (d) $\{+,+,-,-\}$ (transient state). In (b) we observe a power law decay for both type of particles with exponent $\sim 1.5$ similar to EW. In (c) for type 1 particle, we observe a power law decay with decay exponent $\sim 2.25$ and an aggregate phase.}\label{fig:Cn_twoparticles}
\end{figure*}

\vskip 0.2cm
\noindent
{\em Cluster size distribution.} The cluster size distribution $\C(n)$ as a function of $n$ for the four distinct sets are shown in Fig.~\ref{fig:Cn_twoparticles}. 
In the set $\{+,-,-,+\}$ (Fig.~\ref{fig:Cn_twoparticles}(a)), both particle types exhibit similar $\mathcal{C}(n)$ behavior, closely resembling that of diffusive particles. In this case, the first species moves uphill and pushes the interface down, while the second moves downhill and pulls the interface up. $\C(n)$ is exponential with decay similar to that obtained for diffusive particles. The system breaks into micro clusters of the same species of particles (see Fig.~\ref{fig:anime}(a) and Supplemental Material~\cite{supple}). At small length scales, this separation into microclusters leads to anti-correlation in the density-density cross-correlation (shown later). At larger length scales, the two species appear completely mixed.   

In the set $\{+,+,-,+\}$ (Fig.~\ref{fig:Cn_twoparticles}(b)), both particles again exhibit similar $\mathcal{C}(n)$ behavior. In this case, the cluster size distribution shows a power law behaviour. It is similar to the situation where activity is set to zero (``passive" particles). Both species are moving downhill and, therefore, can be called ``valley-seeking". However, while one tries to pull the interface up locally, the other tries to push it down. Stability is achieved when the number of the two species at a site is the same on average (see Fig.~\ref{fig:anime}(b) and Supplemental Material~\cite{supple}). This leads to a thorough mixing of both species at all length scales. The cross-correlation shows highly correlated behaviour at small length scales, unlike the previous scenario (shown later). Note that particle overlap in a given site is allowed in our system. Therefore, even when the system is thoroughly mixed, there could be large clusters of individual species, as evident in the cluster size distribution. 

In the set $\{-,-,-,+\}$ (Fig.~\ref{fig:Cn_twoparticles}(c)), the two species show very different cluster size distributions. Here, the species that moves uphill and pulls the interface up shows a power law decay and a large cluster similar to conserved mass aggregation models. We will discuss this similarity in detail later. On the other hand, the species that moves downhill and pulls the interface up shows an exponential distribution with a decay similar to the purely diffusive case. Since both species pull the interface up, the species that moves uphill favourably pile up, creating a smooth slope (see Fig.~\ref{fig:anime}(c) and Supplemental Material~\cite{supple}). The cluster tries to smooth the rough parts of the interface, causing the cluster to move in a specific direction. However the other species tries to make the interface rough. When the larger cluster sees a roughness profile on both sides, it can flip its orientation and, hence, its direction of movement.

\begin{figure}[t]
        \includegraphics[width=1.0\linewidth]{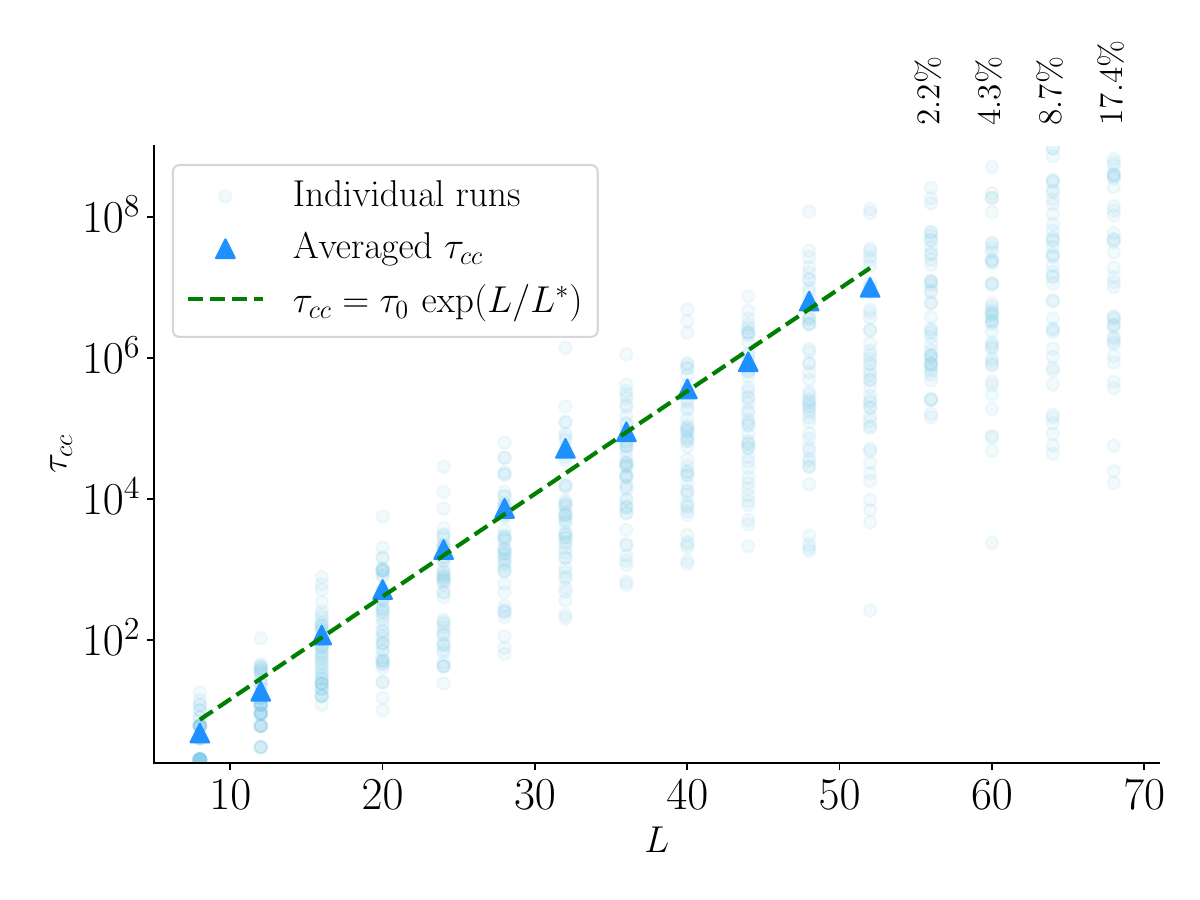}
        \caption{Time for complete clustering ($\tau_{cc}$) vs. membrane length ($L$) for the $\{+,+,-,-\}$ configuration. Clustering time for individual simulation runs is shown as translucent points. Filled triangles mark the average clustering time. For higher $L$ values, individual runs do not reach a clustered state within the simulation cutoff ($10^9$ time steps). The percentage of such failed events is shown. Fitting the average clustering time for smaller $L$ values shows an exponential dependence with $\tau_0 = 0.5; L^* \approx 3.$}~\label{fig:completeclusteringtime}
\end{figure}

In the set $\{+,+,-,-\}$ (Fig.~\ref{fig:Cn_twoparticles}(d)), the two species again show similar behavior of $\C(n)$. Since reaching a steady state is challenging due to extra slow coarsening, we instead present a transient state of $\C(n)$; the system appears to quench before achieving maximal separation. Note that, the behavior is very different from all the other cases described above. For this set, the two types of particles affect the interface in opposite directions. Particle 1 goes downhill and pushes the interface downward, whereas particle 2 goes uphill and pulls the interface up. In this way, particle 1 is ``valley-seeking," and particle 2 is ``hill-seeking." This preferential movement of particles promotes a locked-in conformation of the interface, with the hill seekers residing on the hills and the valley seekers residing in the valleys (see Fig.~\ref{fig:anime}(d) and Supplementary Material~\cite{supple}). Nonetheless, given sufficient time, clusters below a certain size threshold will disappear, as all particles eventually coalesce into two large clusters corresponding to particles 1 and 2. In Fig~\ref{fig:completeclusteringtime}, the time needed for complete clustering ($\tau_{cc}$) is plotted against length $L$. $\tau_{cc}$ exponentially increases with system size, restricting us from reaching the steady state for large $L$. In the transient regime, although the larger clusters remain relatively stable, they occasionally and momentarily fragment into two smaller clusters (due to the Monte-Carlo update), which contributes to the left side of the distribution.

\vskip 0.2cm
\noindent
{\em Comparison to conserved mass aggregation models (CMAM).} The emergent behaviour of type-1 particles in the $\{-,-,-,+\}$ configuration as shown in Fig.~\ref{fig:Cn_twoparticles}(c) exhibits notable similarities to CMAM~\cite{evans2006canonical,majumdar1998nonequilibrium,majumdar2005nature}. To compare the two, we first make the following observation in relation to our system: $(i)$ The cluster moves away, leaving behind a smooth slope (straight section) on one side as its trail $(ii)$ This smooth slope is constantly eroded by `type-2' particles.  $(iii)$ When the opposite trail cluster meets, both sides have a slope, and the cluster remains stationary. The erosion of these slopes continues until one side becomes rough, leading to further dynamics/movement of the cluster. $(iv)$ When the cluster has a smooth slope on only one side, type-2 particles traverse the slope in a manner analogous to objects gliding through a chute.

\vspace{0.2cm}\noindent
{\em Analogy with Chipping:} During the particle update process, if a particle within a cluster is selected, it moves either to the left or right. If the particle's new position lies in a valley, it may chip off from the cluster in the next update step. In this way, a form of chipping occurs within our system. However, this is an emergent effect and cannot be tuned systematically.

\vspace{0.2cm}\noindent{\em Analogy with Aggregation:} When two clusters come into close proximity, they can merge to form a single, stable cluster. This behaviour resembles the aggregation process observed in CMAM, where clusters combine upon contact to create a new, stable structure.

\vspace{0.4cm}\noindent
{\em Non-Markovian Movement of Clusters:}
The history of cluster movement is encoded in the straight sections (smooth slope) formed during its motion. These straight sections are eroded by type-2 particles over time. The straight section restricts the movement of the cluster on that side, causing particles to move only on the other side. The particles hop back and forth between the hill where the cluster currently resides and a neighbouring site, with no other options until a valley appears nearby. Type-1 particles convert this valley into a hill, causing the cluster particles to centre around the newly formed hill. The earlier hill has now formed part of the straight section. This entire process, therefore, guides the motion of a cluster, the straight section acting as a memory trail for its movement. In contrast, in CMAM, a whole cluster diffuses.

The behavior of type-1 particles changes significantly in the presence of type-2 particles. The cluster remains mostly stationary in a single-particle system with stable, straight slopes on both sides. However, type-2 particles erode one of these slopes in a two-particle system, allowing the rough interface to move closer to the stable cluster. Without type-2 particles, the rough interface struggles to approach the cluster, except for occasional small fluctuations that cause minor lateral shifts and slight interface advancement.

\begin{figure}[t]
        \includegraphics[width=1.0\linewidth]{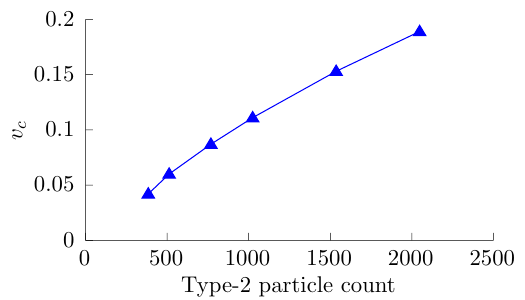}
        \caption{Average speed of cluster ($v_c$) against total number of type-2 particles present, for $\{-,-,-,+\}$. Type-1 particle count: $2048$.}~\label{fig:speed}
\end{figure}

We can extract a characteristic velocity for the motion in the initial part. When we have a large single cluster, we track its average position over time to calculate its average speed. In practice, small pieces may briefly break away from the main cluster, but for our analysis, we disregard this and assume the cluster remains intact throughout the entire observation period. While the cluster may change its orientation, we ensured its direction of movement stayed consistent during the observation time frame. We have plotted this against the number of type-2 particles in Fig \ref{fig:speed}. As observed and expected, the number of type 2 particles significantly impacts the cluster speed.

\begin{figure}[!htbp]
        \includegraphics[width=1.0\linewidth]{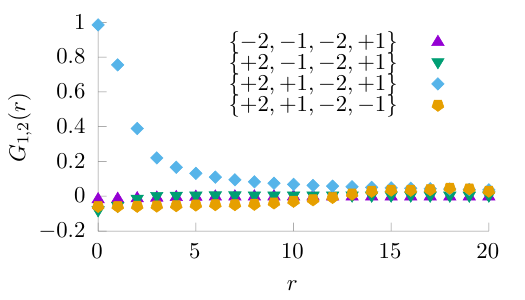}
        \caption{Density-density cross-correlation function for type-1 and type-2 particles. Plots depict all four distinct particle configurations defined by the $\{\beta_1,\gamma_1,\beta_2,\gamma_2\}$ parameter set, elucidating the interaction-driven influence on spatial distribution.}~\label{fig:correlationtwo}
\end{figure}

\vskip 0.2cm
\noindent
{\em Density-density cross-correlation.} We calculate the density-density cross-correlation $G_{1,2}(r)$, between the two species of particles. For the case of set $\{+,-,-,+\}$, as discussed in the previous section, the separation of the system into tiny microclusters of the same species of particles leads to anti-correlation behaviour at small scales, as seen in Fig.~\ref{fig:correlationtwo}. Similar type of results are observed for the cases $\{-,-,-,+\}$ and $\{+,+,-,-\}$. However, for case $\{+,+,-,+\}$, there is a thorough mixing of particles and $G_{1,2}(r)$ shows very strong correlation. Therefore, the cross-correlation behaviour can differentiate between some configurations of the two species of particles. However, for others, the cluster size distribution gives a much clearer understanding.

\vskip 0.2cm
\noindent
The two-species system exhibits diverse behaviours depending on the interaction between activity ($\beta$) and gravity ($\gamma$). Some configurations promote anti-clustering, leading to distinct microclusters, while others result in a uniform mixing of particles. In certain cases, spatial separation occurs, with one species accumulating in valleys and the other on hills. In some, a big cluster forms, collecting all the particles. These variations highlight how tuning the interaction parameters governs particle patterns on the interface.

\begin{comment}
The function $G_{1,2}(r)$, which represents the normalised cross-correlation between two types of particles, $1-\text{type}$ and $2-\text{type}$, as a function of distance $r$, is defined as
\begin{widetext}   
\begin{equation}
    G_{1,2}(r)=
    \frac{\langle s_{1}(i)s_{2}(i+r)\rangle-\langle s_{1}(i)\rangle\langle s_{2}(i+r)\rangle}{\sqrt{\langle s_1(i)^2\rangle-\langle s_1(i)\rangle^2}\sqrt{\langle s_2(i+r)^2\rangle-\langle s_2(i+r)\rangle^2}},
~\label{eq:cor}
\end{equation}
\end{widetext}
where $i$ represents the lattice index and $r$ is in the lattice distance unit. 
\end{comment}

\vskip 0.2cm
\noindent
{\em Comparison to the LH model.} The observed behaviour is very different from the LH model. Coupled with the fact that we do not have hardcore constraints as in the LH model, we also observed different organisations of the two species of particles. In the LH model, for the strong phase separation (SPS) case, the two species (L and H) separate completely, forming a fixed interface shape. Our system exhibits a similar behavior in the  $\{+,+,-,-\}$  configuration. Eventually, the particle species will fully separate, but due to extremely slow coarsening dynamics, the system remains trapped in a prolonged transient (lock-in) state. This transient lock-in state delays reaching the steady state, where complete phase separation is ultimately expected. The finite current with phase separation (FPS) and the fluctuation-dominated phase ordering (FDPO) phase in the LH model occurs when both particles push the interface down, but in the first case with different rates and in the second with the same rate. The particle update for both FPS and FDPO is similar to the $\{-,-,-,+\}$ case in our model and is exactly the same with FDPO for the specific choice of parameters $\beta_1 = \beta_2$. However, the steady-state particle organisation observed in our system is very different. One type of particle forms a huge cluster, showing movement and slope reorientation. The infinitesimal current with phase separation (IPS) is one where H particles tend to push the landscape downward. In contrast, the L particles do not impart any local bias to the landscape dynamics. This is similar to when we have only pusher particles, with the other particles becoming inert. 

\section{Discussions}

We have discussed the effect of active coupling between an interface and multiple species of particles, which move on it stochastically and can deform it locally, acting as ``pushers" or ``pullers". We have shown that the collective effect of a single species of ``puller" particles shows correlated and anti-correlated behaviour depending on two parameters influencing the particle-interface coupling. We have further shown how two different species of particles which sense the local curvature of the interface and affect the underlying interface differently can lead to various organisations of the species on the interface. The particles show separation into small clusters, formation of large clusters, lock-in conformations and uniform mixing. 

Although this paper primarily investigated particle organisation, the key finding is that a cooperative effect of different active particles can exert various forces on the interface, including pushing, pulling, or neutral (passive). The interface behaviour can be significantly altered by manipulating particle activity or introducing new particles. For instance, in the two cases $\{+,-,-,+\}$ and $\{+,+,-,+\}$, the first species is the pusher, while the second is the puller. In both cases, the interface would grow if there was a single species (say, puller). However, when a pusher species is added, it halts interface growth. Specifically, for the case $\{+,+,-,+\}$ we get a passive interface. In the other case, $\{-,-,-,+\}$ particles of type-1 tend to cluster and move collectively in a specific direction, while the type-2 particles effectively gather type-1 particles and guide their movement.

Experiments have revealed that a coupling between the membrane, curved membrane components and the cell cytoskeleton can lead to organised dynamical structures in the cell membrane. Varied curved membrane components and the cytoskeletal forces can affect local membrane curvature, causing it to bend outwards or inwards. Our work, which builds on previous lattice model studies of particle-interface interactions, shows that cooperative effects between multiple particle species can lead to a dynamic organisation on the cell surface and have consequences on membrane deformations. 

The study could be relevant for technological applications in thin film growth. During thin solid film growth using molecular beam epitaxy or vapour deposition, more than one particle is often deposited to tune surface properties. The resultant surface structure and particle organisation reflect the history of the growth process~\cite{drossel2000phase}.  
Our work of adding particles of different types that modify the surface locally suggests that one can control the surface properties by strategically adding or removing specific particle types on a surface. Adjusting the activity levels can lead to selective dispersal of particle types, enabling precise manipulation of surface movement and geometry. 

\section{Data availability statement}
The data that support the findings of this article are openly available~\cite{DataAvailability}.

%\section{Supplementary Material}
%
%The supplementary material includes four videos that show the time evolution of particle distributions and the interface height. Supplementary Videos I, II, III, and IV correspond to the cases $\{+,-,-,+\}$ $\{+,+,-,+\}$$\{-,-,-,+\}$ and $\{+,+,-,-\}$ respectively.
%
%In each video, the particle distributions for type-1 and type-2 particles are plotted in opposite directions for clarity. Specifically, type-1 particles are shown above the $x$-axis, and type-2 particles are plotted below the $x$-axis. This allows for a direct comparison of the spatial distribution between the two particle types. The interface height is plotted after subtracting the mean membrane height to focus on the local features and deformations of the membrane.

%\bibliographystyle{apsrev4-2}
%\bibliography{references.bib}

\begin{thebibliography}{45}%
\makeatletter
\providecommand \@ifxundefined [1]{%
 \@ifx{#1\undefined}
}%
\providecommand \@ifnum [1]{%
 \ifnum #1\expandafter \@firstoftwo
 \else \expandafter \@secondoftwo
 \fi
}%
\providecommand \@ifx [1]{%
 \ifx #1\expandafter \@firstoftwo
 \else \expandafter \@secondoftwo
 \fi
}%
\providecommand \natexlab [1]{#1}%
\providecommand \enquote  [1]{``#1''}%
\providecommand \bibnamefont  [1]{#1}%
\providecommand \bibfnamefont [1]{#1}%
\providecommand \citenamefont [1]{#1}%
\providecommand \href@noop [0]{\@secondoftwo}%
\providecommand \href [0]{\begingroup \@sanitize@url \@href}%
\providecommand \@href[1]{\@@startlink{#1}\@@href}%
\providecommand \@@href[1]{\endgroup#1\@@endlink}%
\providecommand \@sanitize@url [0]{\catcode `\\12\catcode `\$12\catcode
  `\&12\catcode `\#12\catcode `\^12\catcode `\_12\catcode `\%12\relax}%
\providecommand \@@startlink[1]{}%
\providecommand \@@endlink[0]{}%
\providecommand \url  [0]{\begingroup\@sanitize@url \@url }%
\providecommand \@url [1]{\endgroup\@href {#1}{\urlprefix }}%
\providecommand \urlprefix  [0]{URL }%
\providecommand \Eprint [0]{\href }%
\providecommand \doibase [0]{https://doi.org/}%
\providecommand \selectlanguage [0]{\@gobble}%
\providecommand \bibinfo  [0]{\@secondoftwo}%
\providecommand \bibfield  [0]{\@secondoftwo}%
\providecommand \translation [1]{[#1]}%
\providecommand \BibitemOpen [0]{}%
\providecommand \bibitemStop [0]{}%
\providecommand \bibitemNoStop [0]{.\EOS\space}%
\providecommand \EOS [0]{\spacefactor3000\relax}%
\providecommand \BibitemShut  [1]{\csname bibitem#1\endcsname}%
\let\auto@bib@innerbib\@empty
%</preamble>
\bibitem [{\citenamefont {Jarsch}\ \emph {et~al.}(2016)\citenamefont {Jarsch},
  \citenamefont {Daste},\ and\ \citenamefont {Gallop}}]{jarsch2016membrane}%
  \BibitemOpen
  \bibfield  {author} {\bibinfo {author} {\bibfnamefont {I.~K.}\ \bibnamefont
  {Jarsch}}, \bibinfo {author} {\bibfnamefont {F.}~\bibnamefont {Daste}},\ and\
  \bibinfo {author} {\bibfnamefont {J.~L.}\ \bibnamefont {Gallop}},\
  }\href@noop {} {\bibfield  {journal} {\bibinfo  {journal} {Journal of Cell
  Biology}\ }\textbf {\bibinfo {volume} {214}},\ \bibinfo {pages} {375}
  (\bibinfo {year} {2016})}\BibitemShut {NoStop}%
\bibitem [{\citenamefont {McMahon}\ and\ \citenamefont
  {Gallop}(2005)}]{mcmahon2005membrane}%
  \BibitemOpen
  \bibfield  {author} {\bibinfo {author} {\bibfnamefont {H.~T.}\ \bibnamefont
  {McMahon}}\ and\ \bibinfo {author} {\bibfnamefont {J.~L.}\ \bibnamefont
  {Gallop}},\ }\href@noop {} {\bibfield  {journal} {\bibinfo  {journal}
  {Nature}\ }\textbf {\bibinfo {volume} {438}},\ \bibinfo {pages} {590}
  (\bibinfo {year} {2005})}\BibitemShut {NoStop}%
\bibitem [{\citenamefont {McMahon}\ and\ \citenamefont
  {Boucrot}(2015)}]{mcmahon2015membrane}%
  \BibitemOpen
  \bibfield  {author} {\bibinfo {author} {\bibfnamefont {H.~T.}\ \bibnamefont
  {McMahon}}\ and\ \bibinfo {author} {\bibfnamefont {E.}~\bibnamefont
  {Boucrot}},\ }\href@noop {} {\bibfield  {journal} {\bibinfo  {journal}
  {Journal of cell science}\ }\textbf {\bibinfo {volume} {128}},\ \bibinfo
  {pages} {1065} (\bibinfo {year} {2015})}\BibitemShut {NoStop}%
\bibitem [{\citenamefont {Kozlov}\ \emph {et~al.}(2014)\citenamefont {Kozlov},
  \citenamefont {Campelo}, \citenamefont {Liska}, \citenamefont {Chernomordik},
  \citenamefont {Marrink},\ and\ \citenamefont
  {McMahon}}]{kozlov2014mechanisms}%
  \BibitemOpen
  \bibfield  {author} {\bibinfo {author} {\bibfnamefont {M.~M.}\ \bibnamefont
  {Kozlov}}, \bibinfo {author} {\bibfnamefont {F.}~\bibnamefont {Campelo}},
  \bibinfo {author} {\bibfnamefont {N.}~\bibnamefont {Liska}}, \bibinfo
  {author} {\bibfnamefont {L.~V.}\ \bibnamefont {Chernomordik}}, \bibinfo
  {author} {\bibfnamefont {S.~J.}\ \bibnamefont {Marrink}},\ and\ \bibinfo
  {author} {\bibfnamefont {H.~T.}\ \bibnamefont {McMahon}},\ }\href@noop {}
  {\bibfield  {journal} {\bibinfo  {journal} {Current opinion in cell biology}\
  }\textbf {\bibinfo {volume} {29}},\ \bibinfo {pages} {53} (\bibinfo {year}
  {2014})}\BibitemShut {NoStop}%
\bibitem [{\citenamefont {Zimmerberg}\ and\ \citenamefont
  {Kozlov}(2006)}]{zimmerberg2006proteins}%
  \BibitemOpen
  \bibfield  {author} {\bibinfo {author} {\bibfnamefont {J.}~\bibnamefont
  {Zimmerberg}}\ and\ \bibinfo {author} {\bibfnamefont {M.~M.}\ \bibnamefont
  {Kozlov}},\ }\href@noop {} {\bibfield  {journal} {\bibinfo  {journal} {Nature
  reviews Molecular cell biology}\ }\textbf {\bibinfo {volume} {7}},\ \bibinfo
  {pages} {9} (\bibinfo {year} {2006})}\BibitemShut {NoStop}%
\bibitem [{\citenamefont {Gov}(2018)}]{gov2018guided}%
  \BibitemOpen
  \bibfield  {author} {\bibinfo {author} {\bibfnamefont {N.}~\bibnamefont
  {Gov}},\ }\href@noop {} {\bibfield  {journal} {\bibinfo  {journal}
  {Philosophical Transactions of the Royal Society B: Biological Sciences}\
  }\textbf {\bibinfo {volume} {373}},\ \bibinfo {pages} {20170115} (\bibinfo
  {year} {2018})}\BibitemShut {NoStop}%
\bibitem [{\citenamefont {Ramaswamy}\ \emph {et~al.}(1999)\citenamefont
  {Ramaswamy}, \citenamefont {Toner},\ and\ \citenamefont
  {Prost}}]{ramaswamy1999nonequilibrium}%
  \BibitemOpen
  \bibfield  {author} {\bibinfo {author} {\bibfnamefont {S.}~\bibnamefont
  {Ramaswamy}}, \bibinfo {author} {\bibfnamefont {J.}~\bibnamefont {Toner}},\
  and\ \bibinfo {author} {\bibfnamefont {J.}~\bibnamefont {Prost}},\
  }\href@noop {} {\bibfield  {journal} {\bibinfo  {journal} {Pramana}\ }\textbf
  {\bibinfo {volume} {53}},\ \bibinfo {pages} {237} (\bibinfo {year}
  {1999})}\BibitemShut {NoStop}%
\bibitem [{\citenamefont {Ramaswamy}\ \emph {et~al.}(2000)\citenamefont
  {Ramaswamy}, \citenamefont {Toner},\ and\ \citenamefont
  {Prost}}]{ramaswamy2000nonequilibrium}%
  \BibitemOpen
  \bibfield  {author} {\bibinfo {author} {\bibfnamefont {S.}~\bibnamefont
  {Ramaswamy}}, \bibinfo {author} {\bibfnamefont {J.}~\bibnamefont {Toner}},\
  and\ \bibinfo {author} {\bibfnamefont {J.}~\bibnamefont {Prost}},\
  }\href@noop {} {\bibfield  {journal} {\bibinfo  {journal} {Physical review
  letters}\ }\textbf {\bibinfo {volume} {84}},\ \bibinfo {pages} {3494}
  (\bibinfo {year} {2000})}\BibitemShut {NoStop}%
\bibitem [{\citenamefont {Gov}(2004)}]{gov2004membrane}%
  \BibitemOpen
  \bibfield  {author} {\bibinfo {author} {\bibfnamefont {N.}~\bibnamefont
  {Gov}},\ }\href@noop {} {\bibfield  {journal} {\bibinfo  {journal} {Physical
  review letters}\ }\textbf {\bibinfo {volume} {93}},\ \bibinfo {pages}
  {268104} (\bibinfo {year} {2004})}\BibitemShut {NoStop}%
\bibitem [{\citenamefont {Liu}\ \emph {et~al.}(2010)\citenamefont {Liu},
  \citenamefont {Sun}, \citenamefont {Oster},\ and\ \citenamefont
  {Drubin}}]{liu2010mechanochemical}%
  \BibitemOpen
  \bibfield  {author} {\bibinfo {author} {\bibfnamefont {J.}~\bibnamefont
  {Liu}}, \bibinfo {author} {\bibfnamefont {Y.}~\bibnamefont {Sun}}, \bibinfo
  {author} {\bibfnamefont {G.~F.}\ \bibnamefont {Oster}},\ and\ \bibinfo
  {author} {\bibfnamefont {D.~G.}\ \bibnamefont {Drubin}},\ }\href@noop {}
  {\bibfield  {journal} {\bibinfo  {journal} {Current opinion in cell biology}\
  }\textbf {\bibinfo {volume} {22}},\ \bibinfo {pages} {36} (\bibinfo {year}
  {2010})}\BibitemShut {NoStop}%
\bibitem [{\citenamefont {Baumgart}\ \emph {et~al.}(2011)\citenamefont
  {Baumgart}, \citenamefont {Capraro}, \citenamefont {Zhu},\ and\ \citenamefont
  {Das}}]{baumgart2011thermodynamics}%
  \BibitemOpen
  \bibfield  {author} {\bibinfo {author} {\bibfnamefont {T.}~\bibnamefont
  {Baumgart}}, \bibinfo {author} {\bibfnamefont {B.~R.}\ \bibnamefont
  {Capraro}}, \bibinfo {author} {\bibfnamefont {C.}~\bibnamefont {Zhu}},\ and\
  \bibinfo {author} {\bibfnamefont {S.~L.}\ \bibnamefont {Das}},\ }\href@noop
  {} {\bibfield  {journal} {\bibinfo  {journal} {Annual review of physical
  chemistry}\ }\textbf {\bibinfo {volume} {62}},\ \bibinfo {pages} {483}
  (\bibinfo {year} {2011})}\BibitemShut {NoStop}%
\bibitem [{\citenamefont {Simunovic}\ \emph {et~al.}(2015)\citenamefont
  {Simunovic}, \citenamefont {Voth}, \citenamefont {Callan-Jones},\ and\
  \citenamefont {Bassereau}}]{simunovic2015physics}%
  \BibitemOpen
  \bibfield  {author} {\bibinfo {author} {\bibfnamefont {M.}~\bibnamefont
  {Simunovic}}, \bibinfo {author} {\bibfnamefont {G.~A.}\ \bibnamefont {Voth}},
  \bibinfo {author} {\bibfnamefont {A.}~\bibnamefont {Callan-Jones}},\ and\
  \bibinfo {author} {\bibfnamefont {P.}~\bibnamefont {Bassereau}},\ }\href@noop
  {} {\bibfield  {journal} {\bibinfo  {journal} {Trends in cell biology}\
  }\textbf {\bibinfo {volume} {25}},\ \bibinfo {pages} {780} (\bibinfo {year}
  {2015})}\BibitemShut {NoStop}%
\bibitem [{\citenamefont {Veksler}\ and\ \citenamefont
  {Gov}(2007)}]{veksler2007phase}%
  \BibitemOpen
  \bibfield  {author} {\bibinfo {author} {\bibfnamefont {A.}~\bibnamefont
  {Veksler}}\ and\ \bibinfo {author} {\bibfnamefont {N.~S.}\ \bibnamefont
  {Gov}},\ }\href@noop {} {\bibfield  {journal} {\bibinfo  {journal}
  {Biophysical journal}\ }\textbf {\bibinfo {volume} {93}},\ \bibinfo {pages}
  {3798} (\bibinfo {year} {2007})}\BibitemShut {NoStop}%
\bibitem [{\citenamefont {Gov}\ and\ \citenamefont
  {Gopinathan}(2006)}]{gov2006dynamics}%
  \BibitemOpen
  \bibfield  {author} {\bibinfo {author} {\bibfnamefont {N.~S.}\ \bibnamefont
  {Gov}}\ and\ \bibinfo {author} {\bibfnamefont {A.}~\bibnamefont
  {Gopinathan}},\ }\href@noop {} {\bibfield  {journal} {\bibinfo  {journal}
  {Biophysical journal}\ }\textbf {\bibinfo {volume} {90}},\ \bibinfo {pages}
  {454} (\bibinfo {year} {2006})}\BibitemShut {NoStop}%
\bibitem [{\citenamefont {Sorre}\ \emph {et~al.}(2012)\citenamefont {Sorre},
  \citenamefont {Callan-Jones}, \citenamefont {Manzi}, \citenamefont {Goud},
  \citenamefont {Prost}, \citenamefont {Bassereau},\ and\ \citenamefont
  {Roux}}]{sorre2012nature}%
  \BibitemOpen
  \bibfield  {author} {\bibinfo {author} {\bibfnamefont {B.}~\bibnamefont
  {Sorre}}, \bibinfo {author} {\bibfnamefont {A.}~\bibnamefont {Callan-Jones}},
  \bibinfo {author} {\bibfnamefont {J.}~\bibnamefont {Manzi}}, \bibinfo
  {author} {\bibfnamefont {B.}~\bibnamefont {Goud}}, \bibinfo {author}
  {\bibfnamefont {J.}~\bibnamefont {Prost}}, \bibinfo {author} {\bibfnamefont
  {P.}~\bibnamefont {Bassereau}},\ and\ \bibinfo {author} {\bibfnamefont
  {A.}~\bibnamefont {Roux}},\ }\href@noop {} {\bibfield  {journal} {\bibinfo
  {journal} {Proceedings of the National Academy of Sciences}\ }\textbf
  {\bibinfo {volume} {109}},\ \bibinfo {pages} {173} (\bibinfo {year}
  {2012})}\BibitemShut {NoStop}%
\bibitem [{\citenamefont {Sharma}\ \emph {et~al.}(2004)\citenamefont {Sharma},
  \citenamefont {Varma}, \citenamefont {Sarasij}, \citenamefont {Gousset},
  \citenamefont {Krishnamoorthy}, \citenamefont {Rao}, \citenamefont {Mayor}
  \emph {et~al.}}]{sharma2004nanoscale}%
  \BibitemOpen
  \bibfield  {author} {\bibinfo {author} {\bibfnamefont {P.}~\bibnamefont
  {Sharma}}, \bibinfo {author} {\bibfnamefont {R.}~\bibnamefont {Varma}},
  \bibinfo {author} {\bibfnamefont {R.}~\bibnamefont {Sarasij}}, \bibinfo
  {author} {\bibfnamefont {K.}~\bibnamefont {Gousset}}, \bibinfo {author}
  {\bibfnamefont {G.}~\bibnamefont {Krishnamoorthy}}, \bibinfo {author}
  {\bibfnamefont {M.}~\bibnamefont {Rao}}, \bibinfo {author} {\bibfnamefont
  {S.}~\bibnamefont {Mayor}}, \emph {et~al.},\ }\href@noop {} {\bibfield
  {journal} {\bibinfo  {journal} {Cell}\ }\textbf {\bibinfo {volume} {116}},\
  \bibinfo {pages} {577} (\bibinfo {year} {2004})}\BibitemShut {NoStop}%
\bibitem [{\citenamefont {Goswami}\ \emph {et~al.}(2008)\citenamefont
  {Goswami}, \citenamefont {Gowrishankar}, \citenamefont {Bilgrami},
  \citenamefont {Ghosh}, \citenamefont {Raghupathy}, \citenamefont {Chadda},
  \citenamefont {Vishwakarma}, \citenamefont {Rao},\ and\ \citenamefont
  {Mayor}}]{goswami2008nanoclusters}%
  \BibitemOpen
  \bibfield  {author} {\bibinfo {author} {\bibfnamefont {D.}~\bibnamefont
  {Goswami}}, \bibinfo {author} {\bibfnamefont {K.}~\bibnamefont
  {Gowrishankar}}, \bibinfo {author} {\bibfnamefont {S.}~\bibnamefont
  {Bilgrami}}, \bibinfo {author} {\bibfnamefont {S.}~\bibnamefont {Ghosh}},
  \bibinfo {author} {\bibfnamefont {R.}~\bibnamefont {Raghupathy}}, \bibinfo
  {author} {\bibfnamefont {R.}~\bibnamefont {Chadda}}, \bibinfo {author}
  {\bibfnamefont {R.}~\bibnamefont {Vishwakarma}}, \bibinfo {author}
  {\bibfnamefont {M.}~\bibnamefont {Rao}},\ and\ \bibinfo {author}
  {\bibfnamefont {S.}~\bibnamefont {Mayor}},\ }\href@noop {} {\bibfield
  {journal} {\bibinfo  {journal} {Cell}\ }\textbf {\bibinfo {volume} {135}},\
  \bibinfo {pages} {1085} (\bibinfo {year} {2008})}\BibitemShut {NoStop}%
\bibitem [{\citenamefont {Gowrishankar}\ \emph {et~al.}(2012)\citenamefont
  {Gowrishankar}, \citenamefont {Ghosh}, \citenamefont {Saha}, \citenamefont
  {Rumamol}, \citenamefont {Mayor},\ and\ \citenamefont
  {Rao}}]{gowrishankar2012active}%
  \BibitemOpen
  \bibfield  {author} {\bibinfo {author} {\bibfnamefont {K.}~\bibnamefont
  {Gowrishankar}}, \bibinfo {author} {\bibfnamefont {S.}~\bibnamefont {Ghosh}},
  \bibinfo {author} {\bibfnamefont {S.}~\bibnamefont {Saha}}, \bibinfo {author}
  {\bibfnamefont {C.}~\bibnamefont {Rumamol}}, \bibinfo {author} {\bibfnamefont
  {S.}~\bibnamefont {Mayor}},\ and\ \bibinfo {author} {\bibfnamefont
  {M.}~\bibnamefont {Rao}},\ }\href@noop {} {\bibfield  {journal} {\bibinfo
  {journal} {Cell}\ }\textbf {\bibinfo {volume} {149}},\ \bibinfo {pages}
  {1353} (\bibinfo {year} {2012})}\BibitemShut {NoStop}%
\bibitem [{\citenamefont {K{\"o}ster}\ \emph {et~al.}(2016)\citenamefont
  {K{\"o}ster}, \citenamefont {Husain}, \citenamefont {Iljazi}, \citenamefont
  {Bhat}, \citenamefont {Bieling}, \citenamefont {Mullins}, \citenamefont
  {Rao},\ and\ \citenamefont {Mayor}}]{koster2016actomyosin}%
  \BibitemOpen
  \bibfield  {author} {\bibinfo {author} {\bibfnamefont {D.~V.}\ \bibnamefont
  {K{\"o}ster}}, \bibinfo {author} {\bibfnamefont {K.}~\bibnamefont {Husain}},
  \bibinfo {author} {\bibfnamefont {E.}~\bibnamefont {Iljazi}}, \bibinfo
  {author} {\bibfnamefont {A.}~\bibnamefont {Bhat}}, \bibinfo {author}
  {\bibfnamefont {P.}~\bibnamefont {Bieling}}, \bibinfo {author} {\bibfnamefont
  {R.~D.}\ \bibnamefont {Mullins}}, \bibinfo {author} {\bibfnamefont
  {M.}~\bibnamefont {Rao}},\ and\ \bibinfo {author} {\bibfnamefont
  {S.}~\bibnamefont {Mayor}},\ }\href@noop {} {\bibfield  {journal} {\bibinfo
  {journal} {Proceedings of the National Academy of Sciences}\ }\textbf
  {\bibinfo {volume} {113}},\ \bibinfo {pages} {E1645} (\bibinfo {year}
  {2016})}\BibitemShut {NoStop}%
\bibitem [{\citenamefont {Dawson}\ \emph {et~al.}(2006)\citenamefont {Dawson},
  \citenamefont {Legg},\ and\ \citenamefont {Machesky}}]{dawson2006bar}%
  \BibitemOpen
  \bibfield  {author} {\bibinfo {author} {\bibfnamefont {J.~C.}\ \bibnamefont
  {Dawson}}, \bibinfo {author} {\bibfnamefont {J.~A.}\ \bibnamefont {Legg}},\
  and\ \bibinfo {author} {\bibfnamefont {L.~M.}\ \bibnamefont {Machesky}},\
  }\href@noop {} {\bibfield  {journal} {\bibinfo  {journal} {Trends in cell
  biology}\ }\textbf {\bibinfo {volume} {16}},\ \bibinfo {pages} {493}
  (\bibinfo {year} {2006})}\BibitemShut {NoStop}%
\bibitem [{\citenamefont {Pr{\'e}vost}\ \emph {et~al.}(2015)\citenamefont
  {Pr{\'e}vost}, \citenamefont {Zhao}, \citenamefont {Manzi}, \citenamefont
  {Lemichez}, \citenamefont {Lappalainen}, \citenamefont {Callan-Jones},\ and\
  \citenamefont {Bassereau}}]{prevost2015irsp53}%
  \BibitemOpen
  \bibfield  {author} {\bibinfo {author} {\bibfnamefont {C.}~\bibnamefont
  {Pr{\'e}vost}}, \bibinfo {author} {\bibfnamefont {H.}~\bibnamefont {Zhao}},
  \bibinfo {author} {\bibfnamefont {J.}~\bibnamefont {Manzi}}, \bibinfo
  {author} {\bibfnamefont {E.}~\bibnamefont {Lemichez}}, \bibinfo {author}
  {\bibfnamefont {P.}~\bibnamefont {Lappalainen}}, \bibinfo {author}
  {\bibfnamefont {A.}~\bibnamefont {Callan-Jones}},\ and\ \bibinfo {author}
  {\bibfnamefont {P.}~\bibnamefont {Bassereau}},\ }\href@noop {} {\bibfield
  {journal} {\bibinfo  {journal} {Nature communications}\ }\textbf {\bibinfo
  {volume} {6}},\ \bibinfo {pages} {8529} (\bibinfo {year} {2015})}\BibitemShut
  {NoStop}%
\bibitem [{\citenamefont {Galic}\ \emph {et~al.}(2012)\citenamefont {Galic},
  \citenamefont {Jeong}, \citenamefont {Tsai}, \citenamefont {Joubert},
  \citenamefont {Wu}, \citenamefont {Hahn}, \citenamefont {Cui},\ and\
  \citenamefont {Meyer}}]{galic2012external}%
  \BibitemOpen
  \bibfield  {author} {\bibinfo {author} {\bibfnamefont {M.}~\bibnamefont
  {Galic}}, \bibinfo {author} {\bibfnamefont {S.}~\bibnamefont {Jeong}},
  \bibinfo {author} {\bibfnamefont {F.-C.}\ \bibnamefont {Tsai}}, \bibinfo
  {author} {\bibfnamefont {L.-M.}\ \bibnamefont {Joubert}}, \bibinfo {author}
  {\bibfnamefont {Y.~I.}\ \bibnamefont {Wu}}, \bibinfo {author} {\bibfnamefont
  {K.~M.}\ \bibnamefont {Hahn}}, \bibinfo {author} {\bibfnamefont
  {Y.}~\bibnamefont {Cui}},\ and\ \bibinfo {author} {\bibfnamefont
  {T.}~\bibnamefont {Meyer}},\ }\href@noop {} {\bibfield  {journal} {\bibinfo
  {journal} {Nature cell biology}\ }\textbf {\bibinfo {volume} {14}},\ \bibinfo
  {pages} {874} (\bibinfo {year} {2012})}\BibitemShut {NoStop}%
\bibitem [{\citenamefont {Suetsugu}\ \emph {et~al.}(2014)\citenamefont
  {Suetsugu}, \citenamefont {Kurisu},\ and\ \citenamefont
  {Takenawa}}]{suetsugu2014dynamic}%
  \BibitemOpen
  \bibfield  {author} {\bibinfo {author} {\bibfnamefont {S.}~\bibnamefont
  {Suetsugu}}, \bibinfo {author} {\bibfnamefont {S.}~\bibnamefont {Kurisu}},\
  and\ \bibinfo {author} {\bibfnamefont {T.}~\bibnamefont {Takenawa}},\
  }\href@noop {} {\bibfield  {journal} {\bibinfo  {journal} {Physiological
  reviews}\ }\textbf {\bibinfo {volume} {94}},\ \bibinfo {pages} {1219}
  (\bibinfo {year} {2014})}\BibitemShut {NoStop}%
\bibitem [{\citenamefont {Leibler}(1986)}]{leibler1986curvature}%
  \BibitemOpen
  \bibfield  {author} {\bibinfo {author} {\bibfnamefont {S.}~\bibnamefont
  {Leibler}},\ }\href@noop {} {\bibfield  {journal} {\bibinfo  {journal}
  {Journal de Physique}\ }\textbf {\bibinfo {volume} {47}},\ \bibinfo {pages}
  {507} (\bibinfo {year} {1986})}\BibitemShut {NoStop}%
\bibitem [{\citenamefont {Reynwar}\ \emph {et~al.}(2007)\citenamefont
  {Reynwar}, \citenamefont {Illya}, \citenamefont {Harmandaris}, \citenamefont
  {M{\"u}ller}, \citenamefont {Kremer},\ and\ \citenamefont
  {Deserno}}]{reynwar2007aggregation}%
  \BibitemOpen
  \bibfield  {author} {\bibinfo {author} {\bibfnamefont {B.~J.}\ \bibnamefont
  {Reynwar}}, \bibinfo {author} {\bibfnamefont {G.}~\bibnamefont {Illya}},
  \bibinfo {author} {\bibfnamefont {V.~A.}\ \bibnamefont {Harmandaris}},
  \bibinfo {author} {\bibfnamefont {M.~M.}\ \bibnamefont {M{\"u}ller}},
  \bibinfo {author} {\bibfnamefont {K.}~\bibnamefont {Kremer}},\ and\ \bibinfo
  {author} {\bibfnamefont {M.}~\bibnamefont {Deserno}},\ }\href@noop {}
  {\bibfield  {journal} {\bibinfo  {journal} {Nature}\ }\textbf {\bibinfo
  {volume} {447}},\ \bibinfo {pages} {461} (\bibinfo {year}
  {2007})}\BibitemShut {NoStop}%
\bibitem [{\citenamefont {Shlomovitz}\ and\ \citenamefont
  {Gov}(2007)}]{shlomovitz2007membrane}%
  \BibitemOpen
  \bibfield  {author} {\bibinfo {author} {\bibfnamefont {R.}~\bibnamefont
  {Shlomovitz}}\ and\ \bibinfo {author} {\bibfnamefont {N.}~\bibnamefont
  {Gov}},\ }\href@noop {} {\bibfield  {journal} {\bibinfo  {journal} {Physical
  review letters}\ }\textbf {\bibinfo {volume} {98}},\ \bibinfo {pages}
  {168103} (\bibinfo {year} {2007})}\BibitemShut {NoStop}%
\bibitem [{\citenamefont {Chaudhuri}\ \emph {et~al.}(2011)\citenamefont
  {Chaudhuri}, \citenamefont {Bhattacharya}, \citenamefont {Gowrishankar},
  \citenamefont {Mayor},\ and\ \citenamefont
  {Rao}}]{chaudhuri2011spatiotemporal}%
  \BibitemOpen
  \bibfield  {author} {\bibinfo {author} {\bibfnamefont {A.}~\bibnamefont
  {Chaudhuri}}, \bibinfo {author} {\bibfnamefont {B.}~\bibnamefont
  {Bhattacharya}}, \bibinfo {author} {\bibfnamefont {K.}~\bibnamefont
  {Gowrishankar}}, \bibinfo {author} {\bibfnamefont {S.}~\bibnamefont
  {Mayor}},\ and\ \bibinfo {author} {\bibfnamefont {M.}~\bibnamefont {Rao}},\
  }\href@noop {} {\bibfield  {journal} {\bibinfo  {journal} {Proceedings of the
  National Academy of Sciences}\ }\textbf {\bibinfo {volume} {108}},\ \bibinfo
  {pages} {14825} (\bibinfo {year} {2011})}\BibitemShut {NoStop}%
\bibitem [{\citenamefont {Cagnetta}\ \emph {et~al.}(2018)\citenamefont
  {Cagnetta}, \citenamefont {Evans},\ and\ \citenamefont
  {Marenduzzo}}]{cagnetta2018active}%
  \BibitemOpen
  \bibfield  {author} {\bibinfo {author} {\bibfnamefont {F.}~\bibnamefont
  {Cagnetta}}, \bibinfo {author} {\bibfnamefont {M.}~\bibnamefont {Evans}},\
  and\ \bibinfo {author} {\bibfnamefont {D.}~\bibnamefont {Marenduzzo}},\
  }\href@noop {} {\bibfield  {journal} {\bibinfo  {journal} {Physical Review
  Letters}\ }\textbf {\bibinfo {volume} {120}},\ \bibinfo {pages} {258001}
  (\bibinfo {year} {2018})}\BibitemShut {NoStop}%
\bibitem [{\citenamefont {Cagnetta}\ \emph {et~al.}(2019)\citenamefont
  {Cagnetta}, \citenamefont {Evans},\ and\ \citenamefont
  {Marenduzzo}}]{cagnetta2019statistical}%
  \BibitemOpen
  \bibfield  {author} {\bibinfo {author} {\bibfnamefont {F.}~\bibnamefont
  {Cagnetta}}, \bibinfo {author} {\bibfnamefont {M.~R.}\ \bibnamefont
  {Evans}},\ and\ \bibinfo {author} {\bibfnamefont {D.}~\bibnamefont
  {Marenduzzo}},\ }\href@noop {} {\bibfield  {journal} {\bibinfo  {journal}
  {Physical Review E}\ }\textbf {\bibinfo {volume} {99}},\ \bibinfo {pages}
  {042124} (\bibinfo {year} {2019})}\BibitemShut {NoStop}%
\bibitem [{\citenamefont {Cagnetta}\ \emph {et~al.}(2022)\citenamefont
  {Cagnetta}, \citenamefont {{\v{S}}kult{\'e}ty}, \citenamefont {Evans},\ and\
  \citenamefont {Marenduzzo}}]{cagnetta2022universal}%
  \BibitemOpen
  \bibfield  {author} {\bibinfo {author} {\bibfnamefont {F.}~\bibnamefont
  {Cagnetta}}, \bibinfo {author} {\bibfnamefont {V.}~\bibnamefont
  {{\v{S}}kult{\'e}ty}}, \bibinfo {author} {\bibfnamefont {M.~R.}\ \bibnamefont
  {Evans}},\ and\ \bibinfo {author} {\bibfnamefont {D.}~\bibnamefont
  {Marenduzzo}},\ }\href@noop {} {\bibfield  {journal} {\bibinfo  {journal}
  {Physical Review E}\ }\textbf {\bibinfo {volume} {105}},\ \bibinfo {pages}
  {L012604} (\bibinfo {year} {2022})}\BibitemShut {NoStop}%
\bibitem [{\citenamefont {Das}\ and\ \citenamefont
  {Barma}(2000)}]{das2000particles}%
  \BibitemOpen
  \bibfield  {author} {\bibinfo {author} {\bibfnamefont {D.}~\bibnamefont
  {Das}}\ and\ \bibinfo {author} {\bibfnamefont {M.}~\bibnamefont {Barma}},\
  }\href@noop {} {\bibfield  {journal} {\bibinfo  {journal} {Physical review
  letters}\ }\textbf {\bibinfo {volume} {85}},\ \bibinfo {pages} {1602}
  (\bibinfo {year} {2000})}\BibitemShut {NoStop}%
\bibitem [{\citenamefont {Das}\ \emph {et~al.}(2001)\citenamefont {Das},
  \citenamefont {Barma},\ and\ \citenamefont {Majumdar}}]{das2001fluctuation}%
  \BibitemOpen
  \bibfield  {author} {\bibinfo {author} {\bibfnamefont {D.}~\bibnamefont
  {Das}}, \bibinfo {author} {\bibfnamefont {M.}~\bibnamefont {Barma}},\ and\
  \bibinfo {author} {\bibfnamefont {S.~N.}\ \bibnamefont {Majumdar}},\
  }\href@noop {} {\bibfield  {journal} {\bibinfo  {journal} {Physical Review
  E}\ }\textbf {\bibinfo {volume} {64}},\ \bibinfo {pages} {046126} (\bibinfo
  {year} {2001})}\BibitemShut {NoStop}%
\bibitem [{\citenamefont {Nagar}\ \emph {et~al.}(2005)\citenamefont {Nagar},
  \citenamefont {Barma},\ and\ \citenamefont {Majumdar}}]{nagar2005passive}%
  \BibitemOpen
  \bibfield  {author} {\bibinfo {author} {\bibfnamefont {A.}~\bibnamefont
  {Nagar}}, \bibinfo {author} {\bibfnamefont {M.}~\bibnamefont {Barma}},\ and\
  \bibinfo {author} {\bibfnamefont {S.~N.}\ \bibnamefont {Majumdar}},\
  }\href@noop {} {\bibfield  {journal} {\bibinfo  {journal} {Physical review
  letters}\ }\textbf {\bibinfo {volume} {94}},\ \bibinfo {pages} {240601}
  (\bibinfo {year} {2005})}\BibitemShut {NoStop}%
\bibitem [{\citenamefont {Nagar}\ \emph {et~al.}(2006)\citenamefont {Nagar},
  \citenamefont {Majumdar},\ and\ \citenamefont {Barma}}]{nagar2006strong}%
  \BibitemOpen
  \bibfield  {author} {\bibinfo {author} {\bibfnamefont {A.}~\bibnamefont
  {Nagar}}, \bibinfo {author} {\bibfnamefont {S.~N.}\ \bibnamefont
  {Majumdar}},\ and\ \bibinfo {author} {\bibfnamefont {M.}~\bibnamefont
  {Barma}},\ }\href@noop {} {\bibfield  {journal} {\bibinfo  {journal}
  {Physical Review E—Statistical, Nonlinear, and Soft Matter Physics}\
  }\textbf {\bibinfo {volume} {74}},\ \bibinfo {pages} {021124} (\bibinfo
  {year} {2006})}\BibitemShut {NoStop}%
\bibitem [{\citenamefont {Gopalakrishnan}(2004)}]{gopalakrishnan2004dynamics}%
  \BibitemOpen
  \bibfield  {author} {\bibinfo {author} {\bibfnamefont {M.}~\bibnamefont
  {Gopalakrishnan}},\ }\href@noop {} {\bibfield  {journal} {\bibinfo  {journal}
  {Physical Review E}\ }\textbf {\bibinfo {volume} {69}},\ \bibinfo {pages}
  {011105} (\bibinfo {year} {2004})}\BibitemShut {NoStop}%
\bibitem [{\citenamefont {Bisht}\ and\ \citenamefont
  {Barma}(2019)}]{bisht2019interface}%
  \BibitemOpen
  \bibfield  {author} {\bibinfo {author} {\bibfnamefont {P.}~\bibnamefont
  {Bisht}}\ and\ \bibinfo {author} {\bibfnamefont {M.}~\bibnamefont {Barma}},\
  }\href@noop {} {\bibfield  {journal} {\bibinfo  {journal} {Physical Review
  E}\ }\textbf {\bibinfo {volume} {100}},\ \bibinfo {pages} {052120} (\bibinfo
  {year} {2019})}\BibitemShut {NoStop}%
\bibitem [{\citenamefont {Chakraborty}\ \emph
  {et~al.}(2017{\natexlab{a}})\citenamefont {Chakraborty}, \citenamefont
  {Chatterjee},\ and\ \citenamefont {Barma}}]{chakraborty2017static}%
  \BibitemOpen
  \bibfield  {author} {\bibinfo {author} {\bibfnamefont {S.}~\bibnamefont
  {Chakraborty}}, \bibinfo {author} {\bibfnamefont {S.}~\bibnamefont
  {Chatterjee}},\ and\ \bibinfo {author} {\bibfnamefont {M.}~\bibnamefont
  {Barma}},\ }\href@noop {} {\bibfield  {journal} {\bibinfo  {journal}
  {Physical Review E}\ }\textbf {\bibinfo {volume} {96}},\ \bibinfo {pages}
  {022127} (\bibinfo {year} {2017}{\natexlab{a}})}\BibitemShut {NoStop}%
\bibitem [{\citenamefont {Chakraborty}\ \emph
  {et~al.}(2017{\natexlab{b}})\citenamefont {Chakraborty}, \citenamefont
  {Chatterjee},\ and\ \citenamefont {Barma}}]{chakraborty2017dynamic}%
  \BibitemOpen
  \bibfield  {author} {\bibinfo {author} {\bibfnamefont {S.}~\bibnamefont
  {Chakraborty}}, \bibinfo {author} {\bibfnamefont {S.}~\bibnamefont
  {Chatterjee}},\ and\ \bibinfo {author} {\bibfnamefont {M.}~\bibnamefont
  {Barma}},\ }\href@noop {} {\bibfield  {journal} {\bibinfo  {journal}
  {Physical Review E}\ }\textbf {\bibinfo {volume} {96}},\ \bibinfo {pages}
  {022128} (\bibinfo {year} {2017}{\natexlab{b}})}\BibitemShut {NoStop}%
\bibitem [{\citenamefont {Nagar}(2006)}]{nagar:thesis}%
  \BibitemOpen
  \bibfield  {author} {\bibinfo {author} {\bibfnamefont {A.}~\bibnamefont
  {Nagar}},\ }\emph {\bibinfo {title} {Steady States of Passive Particles
  Sliding on Fluctuating Surfaces}},\ \href@noop {} {Ph.D. thesis},\ \bibinfo
  {school} {Citeseer} (\bibinfo {year} {2006})\BibitemShut {NoStop}%
\bibitem [{sup()}]{supple}%
  \BibitemOpen
  \href@noop {} {\bibinfo {title} {See supplementary material at}},\ \bibinfo
  {howpublished} {\url{http://link.aps.org/supplemental/10.1103/PhysRevE.111.045412}},\ \bibinfo {note} {for the four videos that show
  the time evolution of particle distributions and the interface profile.
  Supplementary videos I, II, III, and IV correspond to the cases
  $\{+,-,-,+\}$, $\{+,+,-,+\}$, $\{-,-,-,+\}$, and $\{+,+,-,-\}$, respectively.
  In each video, type-1 particle distributions are displayed above the $x$-axis
  while type-2 are shown below it. The interface height is plotted after
  subtracting the mean membrane height to emphasize local features and membrane
  deformations.}\BibitemShut {Stop}%
\bibitem [{\citenamefont {Evans}\ \emph {et~al.}(2006)\citenamefont {Evans},
  \citenamefont {Majumdar},\ and\ \citenamefont {Zia}}]{evans2006canonical}%
  \BibitemOpen
  \bibfield  {author} {\bibinfo {author} {\bibfnamefont {M.}~\bibnamefont
  {Evans}}, \bibinfo {author} {\bibfnamefont {S.~N.}\ \bibnamefont
  {Majumdar}},\ and\ \bibinfo {author} {\bibfnamefont {R.}~\bibnamefont
  {Zia}},\ }\href@noop {} {\bibfield  {journal} {\bibinfo  {journal} {Journal
  of Statistical Physics}\ }\textbf {\bibinfo {volume} {123}},\ \bibinfo
  {pages} {357} (\bibinfo {year} {2006})}\BibitemShut {NoStop}%
\bibitem [{\citenamefont {Majumdar}\ \emph {et~al.}(1998)\citenamefont
  {Majumdar}, \citenamefont {Krishnamurthy},\ and\ \citenamefont
  {Barma}}]{majumdar1998nonequilibrium}%
  \BibitemOpen
  \bibfield  {author} {\bibinfo {author} {\bibfnamefont {S.~N.}\ \bibnamefont
  {Majumdar}}, \bibinfo {author} {\bibfnamefont {S.}~\bibnamefont
  {Krishnamurthy}},\ and\ \bibinfo {author} {\bibfnamefont {M.}~\bibnamefont
  {Barma}},\ }\href@noop {} {\bibfield  {journal} {\bibinfo  {journal}
  {Physical review letters}\ }\textbf {\bibinfo {volume} {81}},\ \bibinfo
  {pages} {3691} (\bibinfo {year} {1998})}\BibitemShut {NoStop}%
\bibitem [{\citenamefont {Majumdar}\ \emph {et~al.}(2005)\citenamefont
  {Majumdar}, \citenamefont {Evans},\ and\ \citenamefont
  {Zia}}]{majumdar2005nature}%
  \BibitemOpen
  \bibfield  {author} {\bibinfo {author} {\bibfnamefont {S.~N.}\ \bibnamefont
  {Majumdar}}, \bibinfo {author} {\bibfnamefont {M.}~\bibnamefont {Evans}},\
  and\ \bibinfo {author} {\bibfnamefont {R.~K.}\ \bibnamefont {Zia}},\
  }\href@noop {} {\bibfield  {journal} {\bibinfo  {journal} {Physical review
  letters}\ }\textbf {\bibinfo {volume} {94}},\ \bibinfo {pages} {180601}
  (\bibinfo {year} {2005})}\BibitemShut {NoStop}%
\bibitem [{\citenamefont {Drossel}\ and\ \citenamefont
  {Kardar}(2000)}]{drossel2000phase}%
  \BibitemOpen
  \bibfield  {author} {\bibinfo {author} {\bibfnamefont {B.}~\bibnamefont
  {Drossel}}\ and\ \bibinfo {author} {\bibfnamefont {M.}~\bibnamefont
  {Kardar}},\ }\href@noop {} {\bibfield  {journal} {\bibinfo  {journal}
  {Physical Review Letters}\ }\textbf {\bibinfo {volume} {85}},\ \bibinfo
  {pages} {614} (\bibinfo {year} {2000})}\BibitemShut {NoStop}%
\bibitem [{\citenamefont {Grover}\ \emph {et~al.}()\citenamefont {Grover},
  \citenamefont {Kapri},\ and\ \citenamefont {Chaudhuri}}]{DataAvailability}%
  \BibitemOpen
  \bibfield  {author} {\bibinfo {author} {\bibfnamefont {L.}~\bibnamefont
  {Grover}}, \bibinfo {author} {\bibfnamefont {R.}~\bibnamefont {Kapri}},\ and\
  \bibinfo {author} {\bibfnamefont {A.}~\bibnamefont {Chaudhuri}},\ }\href@noop
  {} {\bibinfo {title} {The data that support the findings of this article are
  openly available at:}}\ \bibinfo {note}
  {https://github.com/tolovegrover/SpatialOrganisationData}\BibitemShut
  {NoStop}%
\end{thebibliography}

%

\end{document}